\documentclass[conference]{IEEEtran}
\usepackage{cite}
\usepackage{amsmath,amssymb,amsfonts}
\usepackage{algorithmic}
\usepackage{subfig}
\usepackage{graphicx}
\usepackage{textcomp}
\usepackage{fancyhdr}
\usepackage[hyphens]{url}
\usepackage[normalem]{ulem}
\usepackage[hyphens]{url}
\usepackage{flushend}

\usepackage{shapepar}
\usepackage{comment}
\usepackage[section]{placeins}
\usepackage[ruled,vlined,linesnumbered]{algorithm2e}
\usepackage[bookmarks=true,breaklinks=true,letterpaper=true,colorlinks,linkcolor=black,citecolor=blue,urlcolor=black]{hyperref}
\usepackage{tikz}
\usepackage{xcolor}
\usepackage{soul}
\usepackage{xspace}
\usepackage[nodisplayskipstretch]{setspace}
\usepackage[normalem]{ulem}
\usepackage{caption}
\usepackage{comment}
\def\BibTeX{{\rm B\kern-.05em{\sc i\kern-.025em b}\kern-.08em
    T\kern-.1667em\lower.7ex\hbox{E}\kern-.125emX}}

\newcommand{\iscasubmissionnumber}{574}
\newcommand{\etal}{\textit{et al.}}

\newcommand{\nael}[1]{\textcolor{blue}{Nael: \em #1}}

\renewcommand{\nael}[1]{}

\title{Spy in the GPU-box: Covert and Side Channel Attacks on Multi-GPU Systems} 
\author{}
\makeatletter
\newcommand{\linebreakand}{%
  \end{@IEEEauthorhalign}
  \hfill\mbox{}\par
  \mbox{}\hfill\begin{@IEEEauthorhalign}
}
\makeatother

\author{\IEEEauthorblockN{Sankha Baran Dutta}
\IEEEauthorblockA{\textit{Pacific Northwest National Laboratory} \\
Richland, WA, USA\\
sankha.b.dutta@pnnl.gov}
\and
\IEEEauthorblockN{Hoda Naghibijouybari}
\IEEEauthorblockA{\textit{Department of Computer Science} \\
\textit{Binghamton University}\\
Binghamton, New York, USA \\
hnaghibi@binghamton.edu}
\and
\IEEEauthorblockN{Arjun Gupta}
\IEEEauthorblockA{\textit{Independent Contributor} \\
arjun.gupta@gmail.com}
 \linebreakand 
\IEEEauthorblockN{Nael Abu-Ghazaleh}
\IEEEauthorblockA{\textit{CSE and ECE Departments} \\
\textit{University of California, Riverside}\\
Riverside, California, USA \\
nael@cs.ucr.edu}

\and

\IEEEauthorblockN{Andres Marquez}
\IEEEauthorblockA{\textit{Pacific Northwest National
Laboratory} \\
Richland, WA, USA\\
Andres.Marquez@pnnl.gov}
\and
\IEEEauthorblockN{Kevin Barker}
\IEEEauthorblockA{\textit{Pacific Northwest National
Laboratory} \\
Richland, WA, USA\\
Kevin.Barker@pnnl.gov}
}
\begin{document}
\maketitle
\thispagestyle{plain}
\pagestyle{plain}


\begin{abstract}
The deep learning revolution has been enabled in large part by GPUs, and more recently accelerators, which make it possible to carry out computationally demanding training and inference in acceptable times.  As the size of machine learning networks and workloads continues to increase, multi-GPU machines have emerged as an important platform offered on High Performance Computing and cloud data centers.  As these machines are shared between multiple users, it becomes increasingly important to protect applications against potential attacks.  In this paper, we explore the vulnerability of Nvidia's DGX multi-GPU machines to covert and side channel attacks.  These machines consist of a number of discrete GPUs that are interconnected through a combination of custom interconnect (NVLink) and PCIe connections.  We reverse engineer the cache hierarchy and show that it is possible for an attacker on one GPU to cause contention on the L2 cache of another GPU.  We use this observation to first develop a covert channel attack across two GPUs, achieving the best bandwidth of 3.95 MB/s.  We also develop a prime and probe attack on a remote GPU allowing an attacker to recover the cache hit and miss behavior of another workload.  This basic capability can be used in any number of side channel attacks: we demonstrate a proof of concept attack that fingerprints the application running on the remote GPU, with high accuracy.  Our work establishes for the first time the vulnerability of these machines to microarchitectural attacks, and we hope that it guides future research to improve their security.


\end{abstract}

\section{Introduction}

GPUs have been an important computational platform enabling a variety of data intensive workloads such as deep neural networks, scientific kernels, cryptocurrency mining and many others.  The size of these workloads continue to increase: for example, training of large deep networks often requires both computational and memory resources that far exceed those of a single GPU.  In response to these trends, Multi-GPU platforms have emerged that offer tightly integrated GPUs, enabling applications that span multiple-GPU with unified memory accesses supported by fast communication fabric. For example, the Nvidia DGX series~\cite{dgx_1} offers a number of server class GPUs that are interconnected through a combination of proprietary high bandwidth interconnect (NVLink) and PCIe.  Other GPU manufacturers are also starting to offer similar products; for example, AMD's crossfire allows building relatively inexpensive multi-GPU configurations~\cite{crossfire}.  It is likely that such systems will continue to grow in terms of the performance of the components (GPUs, interconnect and memory) as well as in the number of GPUs that can be supported on each machine.

In this paper, we explore whether multi-GPU machines are vulnerable to both covert and side channel attacks.  Given the importance of workloads that run on these machines, it is important to understand their security properties.  On multi-GPU machines multiple applications may concurrently execute to more effectively use the available resources. Applications generally belong to different mutually untrusting users.  In our threat model, an application either covertly communicates with another (covert channel) or attempts to spy on them (side-channel).  
Covert and side channel attacks have been demonstrated on a variety of CPU microarchitectural structures~\cite{maurice2015c5,saileshwar2021streamline,yaromflush,liu2015last}.  More recently, attacks have been demonstrated on GPUs as well~\cite{hoda_ccs,hoda_micro,jiang-2016,jiang-2017,Nayak-2021}.

Our work demonstrates for the first time that microarchitectural covert and side channel attacks are also dangerous in the context of multi-GPU systems.   Specifically, we first reverse engineer the caches on multi-GPU systems, and discover that they are shared in a Non-Uniform Memory Access (NUMA) configuration: the L2 cache on each GPU caches the data for any memory pages mapped to that GPU's physical memory (even from a remote GPU).  This observation enables us to create contention on remote caches by allocating memory on the target GPU, which is the essential ingredient enabling our covert and side channels.  
Specifically, we develop the first \textbf{microarchitectural covert and side-channel attacks across GPUs in a multi-GPU servers (an Nvidia DGX-1 server)}. In the covert channel attack, a trojan process is located on one GPU transferring secret information to a spy which is located on another GPU. 
In our side channel attacks, the malicious process can monitor the shared L2 cache from a remote GPU and infer secrets about the victim process.   


Cross-GPU attacks offer the attacker a number of advantages compared to prior attacks targeting GPUs.
First, they relieve the attacker from the issue of manipulating the scheduler on a single-GPU to establish co-location of the attacker kernels with the victim (e.g., on the same SM)~\cite{hoda_ccs}. In addition, these attacks also bypass isolation-based defenses such as 
partitioning-based~\cite{GPUGUARD} defense mechanisms that can be enabled for processes running within a single GPU.  Moreover, previous side-channel attacks on a single GPU exploit the aggregate measures of contention on GPUs~\cite{hoda_ccs,jiang-2016,jiang-2017, Wei-2020, Zou-2019, liu-2019}. The attacks that we develop in  this  paper,  are  the  first  Prime+Probe  based timing attacks on  L2  cache  on  GPUs.  Our attacks  extract contention information at the granularity  of  a  single  cache set, providing high-resolution attacks with fine-grained access time  measurements, reducing the noise, and achieving high quality channels. The attacks are conducted entirely from the user level without any special access (e.g.  huge  pages  or  flush  instruction). As a result, we believe this attack model challenges assumptions from prior GPU based attacks and significantly expands our understanding of the threat model in Multi-GPU servers.   


The attacks also require resolving a number of new challenges that are specific to this environment and the userspace only nature of the attack. To develop an attack over the L2 cache we have to reverse engineer the cache architectural details from user space.  Usually system level assistance like using huge pages \cite{irazoqui2015s}\cite{liu2015last} and cache flush instructions provides additional information during reverse engineering, which are unavailable to us.  We first reverse engineer the sharing of the caches discovering that they are configured such that each physical page gets cached at the GPU connected to the memory where it is placed.   Thus, a GPU can remotely access the caches of other GPUs.  We determine the timing characteristics corresponding to access times under different caching scenarios, and use them to develop eviction sets --collections of memory addresses hashing to the same cache set-- both from local and remote GPUs.  
For the covert channel attack, the next challenge is to align the different discovered eviction sets such that the contention is created at the same physical set from both processes.  
Without this alignment, the two sides cannot know which eviction sets to use to cause the contention necessary for the attack. 
 We solve all of the challenges above, enabling a high quality, high bandwidth, prime-and-probe covert channel across GPUs, achieving a bandwidth of 3.95 MBps, with a low error rate of 1.3\%.  Using additional parallelism (e.g., involving additional GPUs) can further improve bandwidth, but we did not explore this in this paper. 
 
 We also explore developing side channel attacks.  The attack relies on recovering the \textit{memorygram} \cite{liu2015last} of the accesses on the cache, and then inferring information about the victim from the distribution of the cache hits and misses.  Specifically, we demonstrate an application/kernel fingerprinting attack where the attacker tries to infer which application is running on a remote GPU.  This attack will be useful as a first step in any other attack to determine where the victim kernels are running.  
 We also demonstrate a simple model extraction attacks which recovers the number of neurons in a hidden layer of a machine learning model ~\cite{DeepSniffer,Wei-2020,hoda_ccs}.

In summary, the contributions of the paper are as follows:

\begin{itemize}
\item We reverse engineer the cache hierarchy and timing properties of  the shared L2 cache in a multi-GPU environment from the user level. The reverse engineering allows us to understand the architectural details of the L2 cache in multi-GPU environment in this modern AI boxes.
\item We identify and overcome challenges that arises in building covert and side channel attacks in a multi-GPU environment.   These challenges include: finding conflict sets for the different cache sets from each process and then aligning the conflict sets across the two processes.

\item We demonstrate the first cross GPU covert channel attack on the shared L2 cache as the attack medium.   With a trojan process on one GPU and a spy process on another, we construct a reliable high-bandwidth covert channel. 

\item We demonstrate side channel attacks  where we construct the memorygram of the accesses of a remote victim.  We use the memorygram in two attacks: (1) fingerprinting applications on the victim GPU; and (2) identifying the number of neurons in a hidden layer of a machine learning model.

\end{itemize}

\section{Background and Threat Model}\label{sec:background}

In this section, we overview the organization of the DGX-1 multi-GPU system from Nvidia which we use as the basis for our attacks and evaluations. We also present the threat model, defining the assumptions we make about the attacker access and capabilities.

 \begin{figure}[t]
    \centering
    \includegraphics[width=\linewidth]{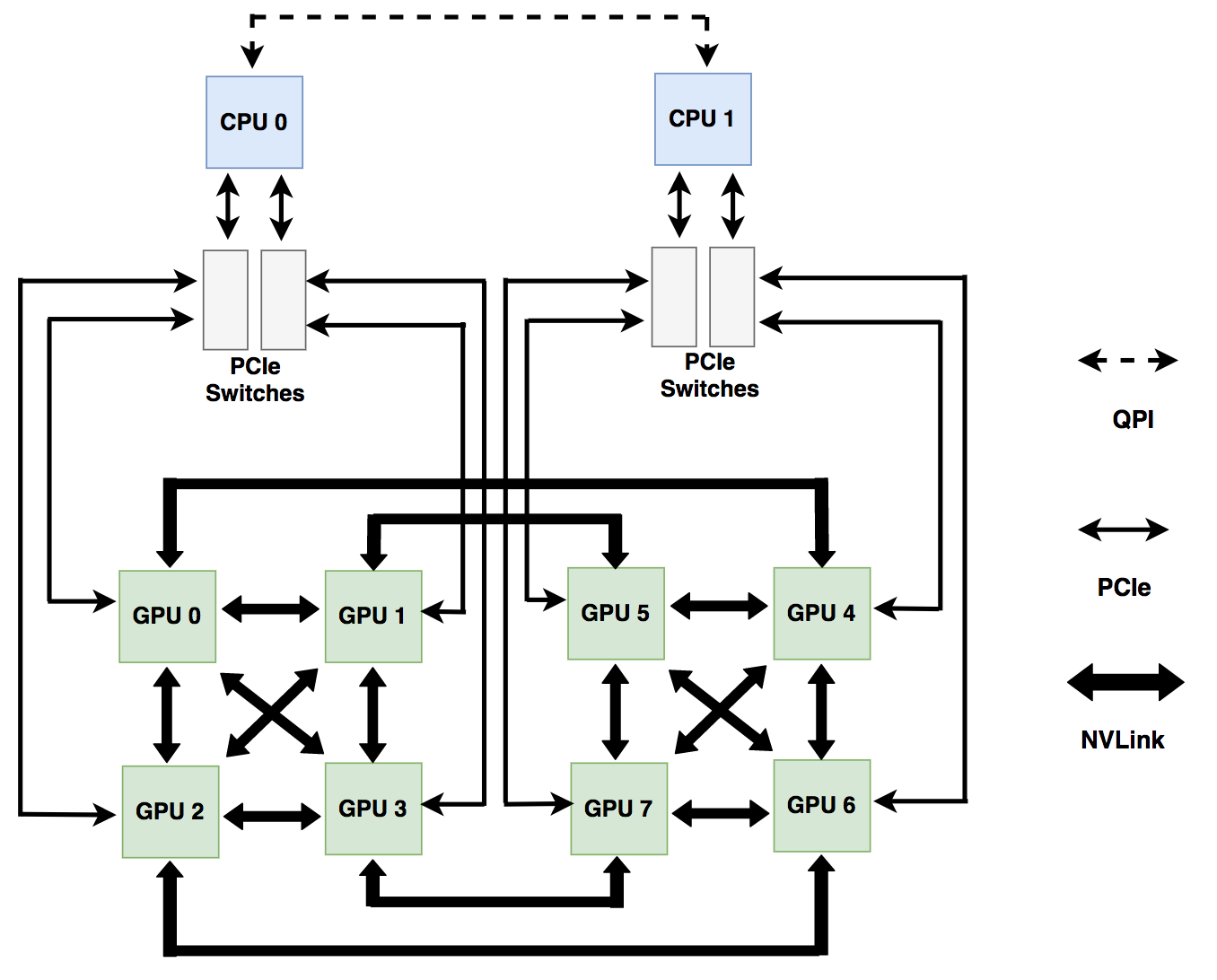}
    \caption{Nvidia DGX-1 topology}
    \label{fig:dgx1}
\end{figure}
\subsection{Multi-GPU Systems}


The demand for high performance computing in deep learning is rapidly growing. As neural networks grow deeper and training data sets become larger, the computational demands to train substantially exceed the capacity of a single GPU, for example requiring weeks to train a large DNN. Therefore, Multi-GPU systems are introduced to provide high throughput and high interconnect bandwidth to maximize neural network training performance.   Multi-GPU systems overcome the memory limitation of a single GPU and offer significant parallelism, and interconnect and memory bandwidth. For example, in 2019, Nvidia used 1,472 V100 interconnected GPUs to bring down the training of a BERT network to 53 minutes~\cite{nvidia-bert}.  Other throughput bound applications could also scale their performance using multiple GPUs.

We develop the attacks in this paper on a Nvidia’s Pascal-based DGX-1 system \cite{dgx_1}. Figure~\ref{fig:dgx1} shows the organization and network topology of this system. DGX-1 box consists of eight Tesla P100 GPUs.  DGX-1 also includes two CPUs (connected through QuickPath Interconnect (QPI)) for boot, storage management, and application coordination. The PCIe links between the GPUs and CPUs enable access to the CPUs’ bulk DRAM memory to enable working set and dataset streaming to and from the GPUs.  The GPUs are connected in a hybrid cube-mesh network topology, with using Nvidia's proprietary NVLink interconnect.  NVLink is an energy-efficient, high-bandwidth interconnect that enables Nvidia GPUs to connect to peer GPUs or other devices within a node at a bidirectional bandwidth of 160 GB/s per GPU: roughly five times that of current PCIe interconnections. 
The GPUs that are connected by NVlink can access each others memories by using Nvidia provided CUDA APIs. 

\begin{figure}
    \centering
    \vspace{-0.2in}
    \includegraphics[width=0.8\linewidth]{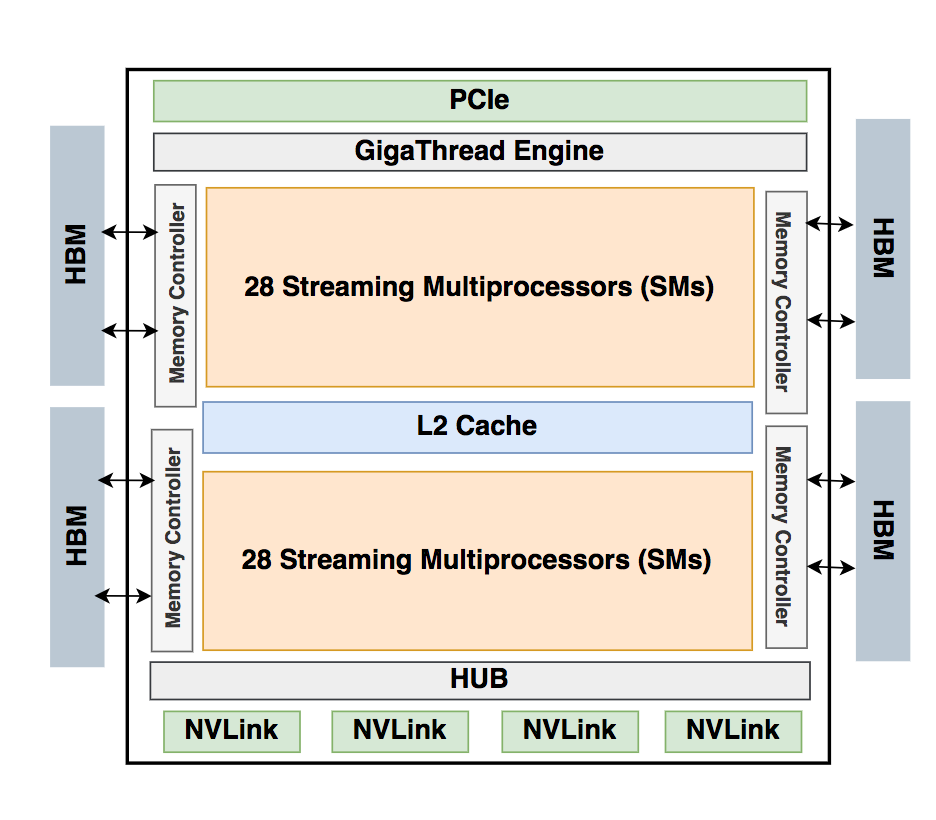}
    \caption{Pascal P100 GPU Architecture}
    \label{fig:p100}
    \vspace{-0.2in}
\end{figure}

GPUs in DGX-1 box are Nvidia's Tesla P100 based on Pascal architecture which is shown in Fig. \ref{fig:p100}. It consists of 56 SMs with a total of 3584 single precision and 1792 double precision units. Each GPU comes with 16 GB of High Bandwidth Memory (HBM2) stacked memory with 732 GB/s of bandwidth. 
There is a private 64KB shared memory per SM and a 4MB L2 cache shared across all SMs. 

\subsection{Threat Model}

In this paper, we develop Prime+Probe based microarchitectural covert and side channel attacks across multiple GPUs on Nvidia's modern GPU servers. Previous microarchitectural attacks were demonstrated on CPU or on a single GPU. However, in our multi-GPU threat model, our attacks span multiple GPUs that are connected via NVLink-V$1$(as shown in Figure~\ref{fig:dgx1}).  The trojan or the victim process is located on a GPU (e.g. GPU 0) and the spy process is located in another GPU (e.g. GPU 1).   Note that if both are located on the same GPU, then prior covert channel attacks on GPUs may be used~\cite{hoda_ccs}; however, we explain later how there are a number of advantages for conducting the attack remotely.

We assume an attacker with normal user access, capable of launching applications on one or more of the GPUs at the same time.   The attacker does not have access to any specialized system support or supervisor privileges. They use experiments to reverse engineer the cache (one time, offline) and to find conflict sets, groups of addresses that hash to the same physical cache set, online as a preliminary step of the attack.  This step is necessary because caches are physically indexed, and sometimes use index hashing, making it difficult to determine the eventual set a virtual address will hash to.  

The attacks are conducted on a DGX-1 box which consists of Pascal based Tesla P100 GPUs programmed using CUDA 10.0.  We expect that with some fine tuning the attacks can be ported to other Multi-GPU systems.  Although we focused on prime-probe attacks exploiting difference in timing between cache hits and misses, we expect that other sources of leakage such as performance counter values can also be used in these attacks~\cite{hoda_ccs,DeepSniffer,Wei-2020}.




\section{Reverse Engineering Cache Organization}\label{sec:Reverse-Eng}

In multi-GPU system, a GPU can access the memory of a remote GPU that is connected via NVLink.  Our attacks are cache based timing attacks. However, the cache hierarchy and its properties is not well documented.  For this reason, we reverse engineer the cache hierarchy and its timing characteristics in this section.

\subsection{Caching organization and timing properties}

In  the  first  set  of  experiments,  our  goal  is  to  understand the overall cache hierarchy as well as the timing properties of different access types (hits vs. misses, local and remote).   The DGX-1  offers  a  uniform  address  space,  and  virtual  pages  can  be allocated  to  physical  pages  that  belongs  in  any  of  the  GPU HBM DRAM memory (i.e., a NUMA organization).  The Pascal  GPUs have  two  levels  of  data  cache,  L1  and  L2.  A  programmer can  bypass  L1  data  caching  by  using  specific  data  loading primitives (specifically, \textunderscore\textunderscore\textit{ldcg()}). However, L2 data caching cannot be bypassed and all data and instructions get cached in L2.

In traditional cache-based timing attacks, an attacker needs to distinguish the cache hit and miss time for different cache levels in order to identify data/cache sets being accessed by the other process (either as part of a covert or side channel attack). In the cross-GPU L2 based timing attack, the attacker needs to understand where data gets cached, and the hit and miss timing properties of both local and remote GPU's caches. Since our attack relies on creating contention between a remote GPU and a local GPU, we developed a microbenchmark to probe for these properties.

We allocate a buffer in the memory of one of the GPUs and use accesses to it to derive both local and remote access times.  To find both the remote and local access time, we first populate the L2 cache with the data from a buffer in DRAM with the stride of 128 bytes which is the L2 cache line size for our Pascal 100 GPU architecture. We use the \textunderscore\textunderscore\textit{ldcg()} load primitive to load the data which allows the data to get cached in the L2 cache only and avoids L1 caching. Each data access is followed by a dummy operation to make sure the access is not optimized out by the compiler. The access time is measured using the clock() function and is recorded in a shared buffer to avoid any contention in the L2 cache as the access path of the shared buffer is separate than the main memory access path. This first cold access shows the DRAM access time. We access the buffer again and measure the access time which represents the L2 cache access time.

\begin{figure}[htbp]
    \centering
    \vspace{-0.1in}
    \includegraphics[width=.95\linewidth]{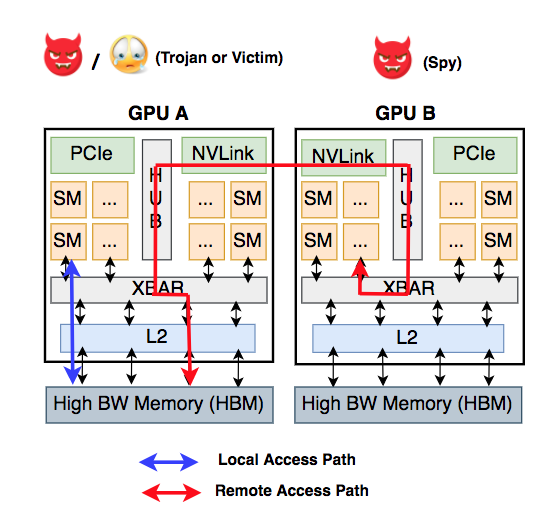}
    \caption{Accessing a remote GPU's memory through NVLink}
    \vspace{-0.2in}
    \label{fig:threatModel_1}
\end{figure}

To measure \textit{remote} L2 hit and miss time, we allocate a buffer on the remote GPU and we use \textit{cudaDeviceEnablePeerAccess} to access the remote GPU's memory. Remote buffer allocation and accessing it does not create any context on the remote GPU, so our two processes have separate contexts that are created on two different GPUs (local and remote).

\begin{figure}[htbp]
    \includegraphics[width=0.8\columnwidth]{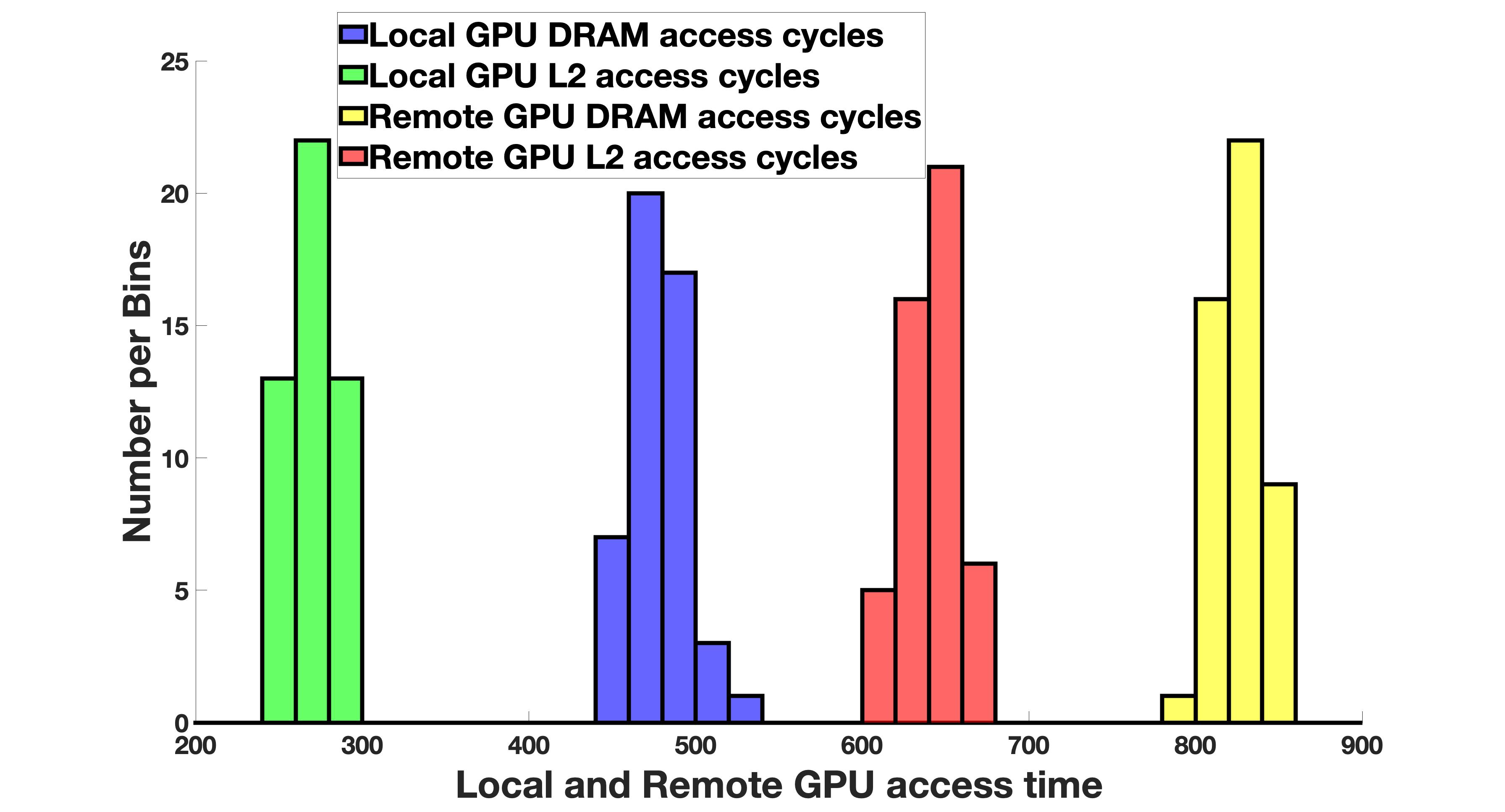}
    \caption{Local and remote GPU access time}
    \label{fig:accessTime}
   \vspace{-0.1in}
\end{figure}

The local and remote GPU L2 and DRAM access time is shown in Fig \ref{fig:accessTime} histogram. We have made 48 accesses in each loop to measure both local and remote GPU. The X-axis specifies the access delay of the data and the Y axis specifies the number of bins in the histogram. As we can see on the figure, there are four clusters of accesses with respect to the timing, varying from just over 250 cycles to over 850 cycles.  When examining the accesses we discover that the fastest accesses (green on the figure) occur to cached accesses to memory pages from the GPU where the memory is allocated.  The next group of accesses correspond to local cache misses: DRAM accesses to the local HBM.  The next two clusters correspond to cache hits on memory that is mapped to a remote GPU, and cache misses to this remote memory respectively.  This experiment indicates that each L2 caches the data for the memory pages mapped to its own memory.  It also provides us with timing thresholds to distinguish between cache hits and misses, to both local and remote GPU caches. We repeated the experiment by selecting different peer-to-peer GPUs connected via NVLink (single-hop) and we have observed similar timing cycle range. NVidia runtime API throws error if the GPUs are not connected via NVLink. 

Thus, we understand the memory access pathways and caching to be as shown in Figure~\ref{fig:threatModel_1}.   When DRAM pages are allocated in the local GPU memory, the data access path is straightforward: the first access is serviced from the local HBM DRAM and subsequent accesses hit the L2 cache on the same device.  On the other hand, when the data is allocated on one GPU and accessed from another, the request is routed through the NVLink connection and the requested cache line is also sent back through NVLink.  Our experiment shows that this data accessed on the remote GPU is cached on the remote GPU, rather on the local L2 GPU.  Of course, caching the data locally, would introduce cache coherence issues since copies of the same data could exist in multiple L2 caches.  
\textbf{In summary, our reverse engineering results demonstrate that an access to the memory of a remote GPU through NVLink is cached on the L2 cache of remote GPU, but not L2 cache of local GPU.} 
We use this shared remote L2 cache in GPU-to-GPU communication to build microarchitectural covert and side channel attacks. 

\subsection{Determining Cache Eviction Sets}\label{sssec:evSets}

To conduct a successful Prime+Probe attack, an attacker needs to find a set of addresses that index into the same cache set.  The number of addresses in this set should at least match the associativity of the cache, such that access to the set replaces current entries in that cache set; such a set is called {\em an eviction set}.   


Although finding conflict sets is a standard component of prime and probe attacks, deriving these sets in our context is somewhat different.  Many (but not all) CPU attacks benefit from additional features such as huge pages which substantially simplify the conflict set derivation.  We also had to to work with the GPU computational model which required care to maintain synchronization.  There are previous studies that explore the L2 cache architecture in the recent GPU architectures.  In general, their assumptions are not compatible with our scenario where the attacker is attempting to find the eviction sets on the fly.  Specifically, Mei \etal \cite{mei2016dissecting} explored different levels of memory hierarchy in GPUs. However, we could not use their attack directly because their reverse engineering process requires storing of all the timer values in shared memory which significantly limits the number of samples we are able to take. Jia \etal \cite{jia2018dissecting} explored different memory levels of Volta and Pascal based architectures. However, they did not provide detailed information about the reverse engineering of L2 architecture (and none for the multi-GPU scenario).  Jain \etal \cite{jain2019fractional} provided detailed information about the L2 reverse engineering as well as the architectural details. However, they modified the driver virtual to physical address translation to force consecutive allocation in the physical address space.  Of course, this property does not hold under our threat model since modifying the driver requires privileged access.  


We use a pointer chase experiment shown in 
Algorithm \ref{alg:evictSetAlgo}. Conceptually, this is similar to traditional prime and probe attacks, but customized to the GPU.  Moreover, since our attack is remote, we are able to substantially accelerate the attack and reduce the noise.  Specifically, all memory used to store measurement values are on the attacker GPU, and therefore they do not generate noise that interferes with the target/remote victim cache.  This enabled us to be more aggressive in deriving the conflict sets.  

The experiment proceeds as follows, a data buffer is allocated and the target index \textit{targetIdx} is chosen in line \ref{alg:evictLine_1} 
The target index is accessed in line \ref{alg:evictLine_2} and the access time is recorded in a time buffer \textit{sharedTimeBuff} allocated in shared memory in line \ref{alg:evictLine_3}. The target address access is followed by accessing other addresses in a loop from line \ref{alg:evictLine_4}. The number of addresses to be accessed is specified by \textit{numOfElements}. The value of \textit{numOfElements} starts with value of one in the first kernel launch.
The access offset is set to 128 bytes which is the cache line size. The number of accesses are increased over subsequent kernel calls which signifies the number of addresses traversed. The target address are accessed again at the end of \textit{for loop}. The second access time of the target address is recorded in another location of the shared buffer. 

After the end of each kernel call these two access times are checked on the host side. The first access time of the target address is the DRAM access time and if the target address resides in the L2 cache then the second access time would be equivalent to the cache access time. However, if the access causes target address to be evicted, then it would be equivalent to the DRAM access time. The target address eviction in this case is caused by the accessing the last address that got accessed.
 This eviction of the target address indicates that the target address 
 and the last address 
 are in the same cache set. Next, to find the rest of the addresses within the same cache set, we remove that last address that caused the eviction from the pointer chase in subsequent kernel calls and continue with the process to find more  addresses that hash to the same set. Also, to find more eviction sets the attacker needs to change the target address and repeats the pointer chase process. To reduce the search space we adopted some optimization methodologies by skipping some address accesses. However, if an eviction is seen in the target address then we revert back and check all those last skipped addresses and determine which exact address causes the eviction of the target address.  This processes can be optimized by observing the data belonging to a page is indexed consecutively in the cache.

\begin{small}
\begin{algorithm}
\SetAlgoLined

basePtr = \&mainBuffer[\textit{targetIdx}];\\\label{alg:evictLine_1}
start = clock();\\
nxtIdx =\textunderscore\textunderscore ldcg(basePtr);\\\label{alg:evictLine_2}
dummy$+$=nxtIdx;\\
end = clock();\\
\textunderscore\textunderscore(\textit{threadfence()});\\
sharedTimeBuff[0] = (end-start);\\\label{alg:evictLine_3}
\textit{dummy Operation}

\For{$i = 0;\ i < numOfElements;\ i = i + 1$}{\label{alg:evictLine_4}
    otherPtr = \&mainBuff[nxtIdx];\\
    nxtIdx = \textunderscore\textunderscore ldcg(otherPtr);\\
    dummy$+$=nxtIdx;\\
    \textunderscore\textunderscore\textit{threadfence()};\\
}

\textit{dummy Operation}

start = clock();\\
nxtIdx =\textunderscore\textunderscore ldcg(basePtr);\\\label{alg:evictLine_5}
dummy$+$=nxtIdx;\\
end = clock();\\
\textunderscore\textunderscore\textit{threadfence()};\\
sharedTimeBuff[1] = (end-start);\\\label{alg:evictLine_6}
\textit{dummy Operation}\\

\caption{Eviction Set Determination Algorithm}
\label{alg:evictSetAlgo}
\end{algorithm}
\end{small}

We also observed that the derived eviction sets remain valid over  application runs as long as the memory allocation size of the process remains unchanged. We also confirmed that the address placement in the cache is independent of the GPU which the kernel is launched on.  These observations allow us to simplify the attack to avoid deriving the conflict sets online in every attack.

The cache line size is 128B and from our eviction set determination experiment, we also learn the associativity of the cache (16).   We repeat the eviction set algorithm \ref{alg:evictSetAlgo} with those recorded addresses only. We observe that the target address is evicted after every 16\textsuperscript{th} address reliably. This implies that there are 16 cache lines in the cache set.  Also, the eviction pattern shows that the replacement policy is LRU (or pseudo-LRU) without randomization since the target address are evicted consistently after 16\textsuperscript{th} address. Table~\ref{tab:cacheArch} summarizes L2 cache parameters and architecture derived from our reverse engineering experiments. 

\begin{table}
\caption{L2 cache architecture}
\centering
\begin{tabular}{ |c|c| } 
\hline
\textbf{Cache Attribute} &  \textbf{Values}\\ 
\hline
L2 cache size & 4MB\\
\hline
Number of Sets   & 2048 \\ 
\hline
Cache line size & 128B\\
\hline
Cache lines per set   & 16\\
\hline
Replacement Policy   & LRU \\  
\hline
\end{tabular}
\label{tab:cacheArch}
\end{table}

\begin{figure}[htbp]
    \centering
    \includegraphics[width=0.8\columnwidth]{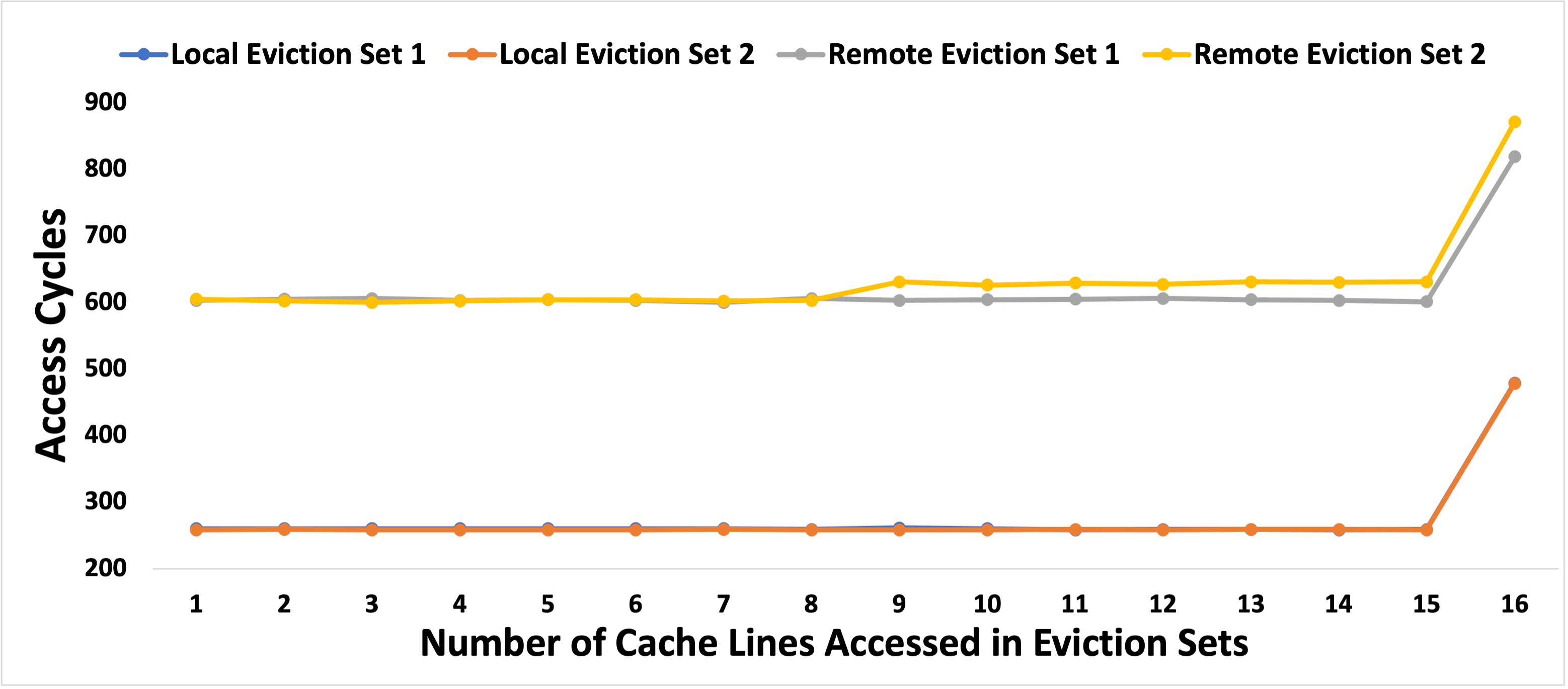}
    \caption{Validating the eviction set determination}
    \label{fig:evSetValid}
    \vspace{-0.1in}
\end{figure}

Figure {\ref{fig:evSetValid}} shows an experiment we conduct for eviction set validation for two derived eviction sets on both the local and remote GPUs. The X-axis is the number of cache lines from the conflict set that have been accessed and the Y-axis is the access time in cycles. We observed that there is an eviction (increase in access time) after every 16\textsuperscript{th} access. This behavior confirms the LRU-based replacement policy with a deterministic replacement for the eviction set access pattern.



\begin{figure}[htbp]
    \centering
    \includegraphics[height=.65\columnwidth,width=.7\columnwidth]{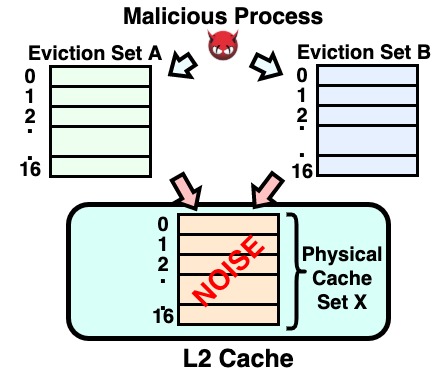}
    \caption{Eviction set aliasing issue}
    \label{fig:evictSetAlias}
    \vspace{-0.15in}
\end{figure}

The GPU L2 cache is physically indexed and the attacker does not have the knowledge of data placement in the cache. As a result, once we discover an eviction set, we are unsure whether it indexes into a new cache set or a previously discovered one.  If we do not ensure that the eviction sets correspond to unique physical sets, this aliasing will result in noise during the attack (Figure \ref{fig:evictSetAlias}). Specifically, there are two eviction sets A and B determined by the malicious process that happen to index to the same physical cache set X, due to the lack of knowledge of the address placement. If there are aliased cache sets within the same process, then during the actual attack phase, the eviction sets would cause interference due to self-eviction leading to the detection of a cache miss and inferring a victim access even when there wasn't one.   Thus, it is important to test each discovered new eviction set against already discovered ones.  If we notice misses when we combine more than 16 addresses from the two sets, we conclude that the two sets correspond to the same physical set and eliminate the newly discovered eviction set from consideration.  

At the conclusion of this process, each process has discovered a collection of eviction sets ideally to cover the full cache.  The reverse engineering results also provide the attacker with timing thresholds to distinguish between cache hits and misses, both on the local GPU as well as the remote GPU.  With this information, we are ready to develop the end to end covert channel attack in the next section.

\section{Covert Channel Attack and Challenges}\label{sec:challenges}

Having established the caching organization and timing characteristics, in this section, we develop a covert channel attack across two GPUs.  Previous GPU-based microarchitectural attacks were demonstrated within a single GPU, and the majority use aggregate measures of contention such as performance counters. Besides establishing this new threat model, the attack has advantages over single GPU attack: it bypasses defenses focused on a single GPU, it reduces the noise, and it avoids having to work around the scheduler to co-locate the two kernels within the GPU so that they can establish contention (e.g., on the same SM~\cite{hoda_micro}). 
The attack is conducted from user level and do not require any system level features such huge pages or flush instructions that are necessary for many attacks. 

In this attack, the Trojan (the transmitter of the covert channel) is located on a local GPU, GPU A, and Spy (the receiver) is located on the remote GPU, GPU B, and accesses the memory of the GPU1 to synchronize and receive information (the opposite is also possible). These two processes communicate covertly over the shared L2 cache of GPU A. First, the spy primes a cache set. To communicate "1", the trojan would access it's own data, evicts the spy's data, and fills up the cache set and to communicate "0", the trojan process does nothing. The spy process keeps probing the same cache set and records the access time. A high access time indicates a miss and interpreted as "1" and a low access time indicates a hit and interpreted as "0".  Although the overall attack process is similar to traditional Prime+Probe attacks, there are several unique challenges that arise due to the platform. 
We  describe these challenges and our approach to overcome them next.

\subsection{Aligning the cache sets}

\begin{figure}[htbp]
    \centering
    \vspace{-.5in}
    \includegraphics[width=\columnwidth,height=.8\columnwidth]{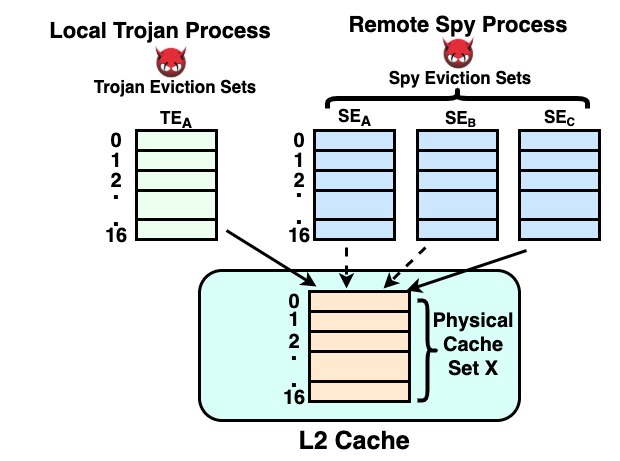}
    \caption{Eviction set alignment among multiple processes}
    \label{fig:evictSetMap}
    \vspace{-0.2in}
\end{figure}

At the stage, the two processes have derived each eviction sets covering the L2 cache.  However, all they are able to determine is that each set hashes to the same physical cache set, but not to which set.  To be able to communicate, the processes have to use eviction set pairs, one in each process, that hash to the same physical cache set.  
We develop a protocol to enable the processes to discover and agree on the sets to use for the signalling and communication as shown in Figure~\ref{fig:evictSetMap}.

Assume again that the trojan process is located on GPU A and the spy process has been launched on GPU B and they are connected via NVLink. Both processes allocated their buffer on GPU A and share the L2 cache on that GPU. In this scenario, the trojan is a local process and the spy is remote. 
In a single run of the malicious applications, one trojan eviction set is checked with another eviction set of the spy process. In Fig \ref{fig:evictSetMap}, we can see a local trojan process eviction set TE\textsubscript{A} launched on GPU A and the remote spy process launched on GPU B have three eviction sets SE\textsubscript{A}, SE\textsubscript{B} and SE\textsubscript{C}. The eviction set of the local trojan process is checked against three eviction sets of the spy process that could be located in the same physical cache set X. The set matching experiment reveals that the trojan eviction set TE\textsubscript{A} is not mapped to the spy eviction set SE\textsubscript{A} and SE\textsubscript{B} shown by dotted arrows. But the trojan eviction set TE\textsubscript{A} is mapped to the spy eviction set SE\textsubscript{C}.

The eviction set mapping kernel is shown in Algorithm \ref{alg:evictSetMap}. Eviction set is accessed from line \ref{alg:mapLn5} - \ref{alg:mapLn13} and the number of access is equivalent to the number of cache lines specified by \textit{numOfCacheLines} (which is 16 in our case). A single eviction set is accessed for \textit{numMainLoop} number of times in \ref{alg:mapLn1}. The actual access of the data takes place in line \ref{alg:mapLn8} and the first index is specified in line \ref{alg:mapLn2} and gets initialized every time within the outer loop. The access takes place in a pointer chase fashion within the inner loop. The access cycles are measured and kept in a register variable \textit{timer1} which accumulates the single access of the eviction set. Another register variable \textit{timer2} in line \ref{alg:mapLn14} accumulates the average access time of a single access of the eviction set. Finally all the accesses over the outer loop are averaged in line \ref{alg:mapLn17}. The kernel algorithm is same for both the trojan and spy processes. The only difference between them is the number of outer loops that decides how many times a cache set would be probed. The trojan process has a faster access compared to the spy process as the memory is local to the trojan process. So the value of \textit{numMainLoop} is much higher for the trojan process compared to the spy process. For our work, we have selected a value of 400000 and 150000 for the local trojan and remote spy process respectively. However, these probing values can be reduced to optimize the execution time of the set mapping process. The main target is to create a visible contention in the L2 cache set and loop boundary controls that contention. 

Note that this particular challenge is required in the covert channel only to communicate between two malicious processes. For side channel, only finding the unique cache sets satisfies the purpose.

\begin{small}
\begin{algorithm}
\SetAlgoLined
\For{$i = 0;\ i < numMainLoop;\ i = i + 1$}{\label{alg:mapLn1}

idxTemp = startIdx;\\\label{alg:mapLn2}
timer1 = 0;\\\label{alg:mapLn3}
dummy1 = 0;\\\label{alg:mapLn4}

\For{$i = 0;\ i < numOfCacheLines;\ i = i + 1$}{\label{alg:mapLn5}

dataPtr = \&mainBuff[idxTemp];\\\label{alg:mapLn6}
start = clock();\\\label{alg:mapLn7}
idxTemp = \textunderscore\textunderscore\textit{ldcg(dataPtr)};\\\label{alg:mapLn8}
dummy1$+$=idxTemp;\\\label{alg:mapLn9}
end = clock();\\\label{alg:mapLn10}
timer1$+$=(end-start);\\\label{alg:mapLn11}
\textunderscore\textunderscore\textit{threadfence()};\\\label{alg:mapLn12}
}\label{alg:mapLn13}
 
timer2$+$=(timer1$/$numOfCacheLines);\\\label{alg:mapLn14}
\textit{dummy operation}\\\label{alg:mapLn15}
}\label{alg:mapLn16}

timeBuffMain = (timer2$/$(numMainLoop));\\\label{alg:mapLn17}

\caption{Eviction set alignment across processes}
\label{alg:evictSetMap}
\end{algorithm}

\end{small}

\subsection{Putting it together: Covert channel attack}

The trojan process (launched on GPU A) allocates the data buffer on the same GPU in step 1 and the spy process gets launched on another GPU (B), but allocates the buffer on the remote GPU A, where the trojan process is launched. The first access of both the trojan and the spy process get the data from the off-chip GPU DRAM and get cached in the L2 cache. The subsequent memory accesses will be serviced from the L2 cache of GPU A.  

\begin{figure}[htbp]
    \centering
    \vspace{-0.1in}
    \includegraphics[width=.9\linewidth,height=.8\linewidth]{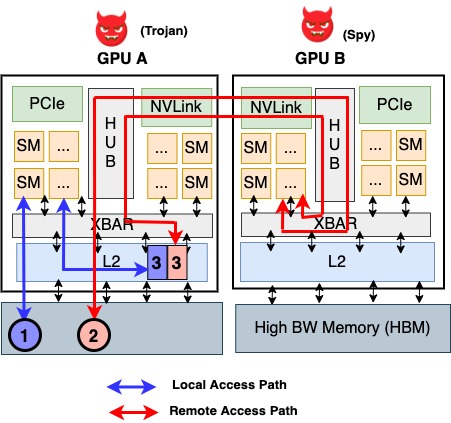}
    \caption{Cross GPU covert channel attack}\nael{Steps 1 to 3 have all been explained--is there nothing else similar to the figures you had in the buddies paper?  How about synchronization?}
    
    \label{fig:covChannFinal}
    \vspace{-0.2in}
\end{figure}

The overall flow of the covert channel attack is shown in Figure \ref{fig:covChannFinal}. The trojan (or sender) is located on GPU A and the spy (receiver) is located on GPU B.  Step 1 and 2 of the attack represent the determination the eviction sets of both processes.  These are followed by the alignment step (Step 3 on the figure) to map the sets on each side to the same physical cache sets, which now enables them to communicate by creating or withholding contention on these sets.   

\nael{This next paragraph is details without intuition first.}
From the previous step of cache set alignment, we have been able to determine the cache sets that are mapped among the malicious processes. This allows us to select the cache sets that would be used during the covert channel communication process. For each cache set we have allocated a thread block that would be launched to a SM in the GPUs. Hence, when the communication takes place over a single cache set, a single thread block on both trojan and spy side would access their own eviction set whose mapping was determined from the set aligning step. We leveraged the GPU parallelism by increasing the number of thread blocks. Each thread block, in both trojan and spy, would access different eviction sets that are already mapped to have a faster communication. 
The trojan thread block consists of a single warp of threads (32 on our machines).  All 32 threads in a thread block of the trojan process are involved in probing the cache set. The 16 addresses referring to the 16 cache lines in the eviction set are accessed through pointer chasing similar to the eviction set determination technique.\nael{What is pointer chase fashion?} The spy process essentially also has 32 threads that are active in the attack; however, we use a significantly higher number of threads (1024) and use the additional threads to help to efficiently save the recorded times from the buffer in shared memory to global memory when it fills.  
Storing the access cycles temporarily on the shared buffer and then copying to the main buffer reduces memory pressure as well as increasing the parallelism during the data copy. 
To send a "1" the trojan process accesses the cache set, replacing the data placed there by the spy, and does nothing to send a "0". 
We have used controlling parameters that control the priming of the cache set while sending a "1", and use computationally heavy dummy instructions (e.g. trigonometric instructions) to wait during transmitting "0" to the spy process. The spy process, however, continuously probes the cache set to receive the data from the trojan process.\nael{This is unclear -- so you do not synchronize?}

\subsection{Covert Channel Evaluation}

\begin{figure*}[t!]
     \centering
     \includegraphics[width=0.9\textwidth]{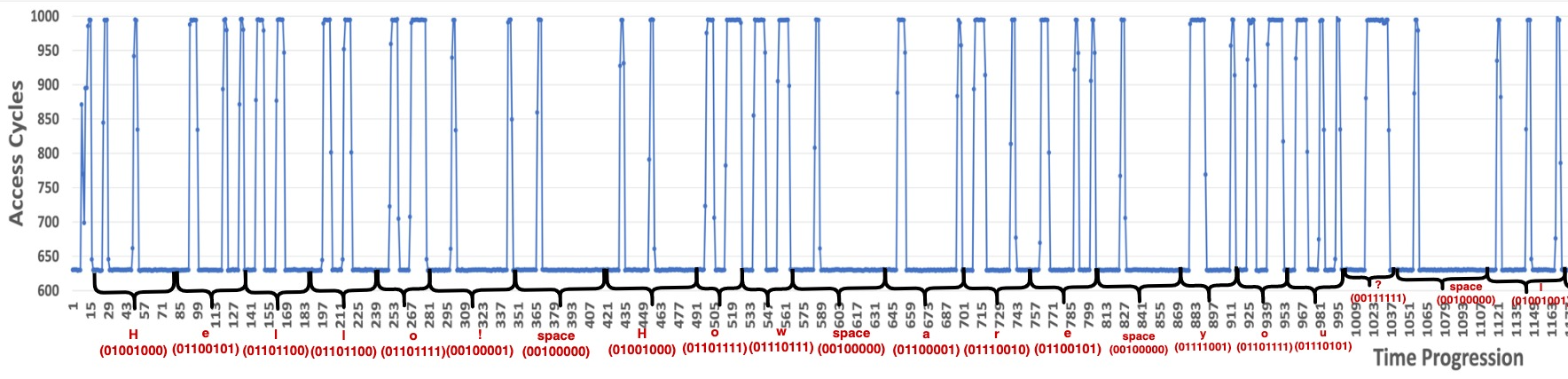}
     \caption{Cross GPU covert message received by spy process}
         \vspace{-0.15in}
\label{fig:covChannMssg_1}
 \end{figure*}

In this section, we evaluate the multi-GPU covert channel attack. All experiment use CUDA 10.0 with Nvidia driver version 410.79. For the covert channel evaluation, we send a long message across the GPUs using L2 cache sets. We send a message of side 1Mb across the covert channel.  We vary the number of cache sets we use in the attack. 
Fig. \ref{fig:covChannMssg_1} shows a demonstration of  the transmission of the first part of a message.  Specifically, the X-axis of the figure is the time progression and the Y axis is the access cycles. The message shows the first line in the text,(\textit{"Hello! How are you? "}) in the long message that have been transferred covertly. The y-axis shows the timed access cycles measured from the remote spy as it accesses the cache set. We observe that the number of cycles is 630 while sending '0' and 950 cycles while communicating '1'. 
To synchronize the communication, as the trojan and the spy processes are located on different GPUs, we tune parameters on the trojan side that controls the cache access frequency to communicate the covert message successfully to the spy side.\nael{Vague and not convincing.  You could easily lose synchronization especially since the 0 and 1 times are different.}

\begin{figure}[htbp]
    \centering
    \vspace{-0.2in}
    \includegraphics[width=0.8\columnwidth]{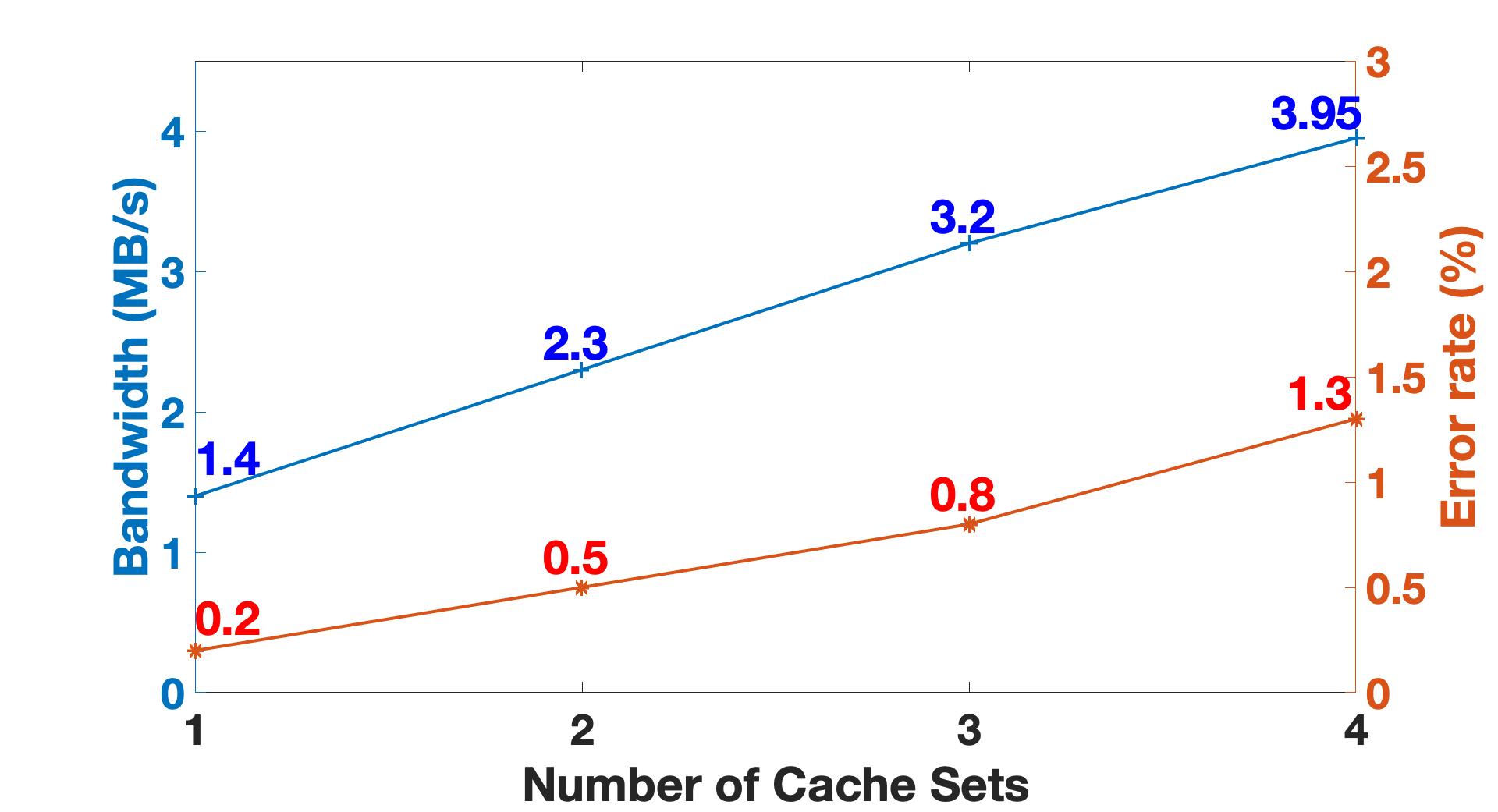}
    \caption{Bandwidth and Error rate in covert channel} 
    \label{fig:covChannBwErr}
    \vspace{-0.1in}
\end{figure}

The bandwidth and the error rate are shown in Fig. \ref{fig:covChannBwErr}. We have measured the bandwidth and error rate as we increase the number of sets used for communication on the x-axis of the figure.  The left y-axis of the figure is the bandwidth corresponding to the blue line in the figure which is displayed in MB$/$s.  Similarly, the right y-axis shows the error rate in percentage corresponding to the red line. 
We measured the bandwidth and the error rate measured over 1000 runs of sending the message from trojan to spy. The bandwidth increases as the number of cache sets increases, since we are able to communicate over multiple cache sets in parallel. However, as the number of cache sets increases, the contention increases among resources such as ports, introducing more variability in the timing, and increasing the error rate increases as well.
\textbf{The highest bandwidth is 3.95 MB$/$s when using 4 cache sets in parallel and with an average error rate of 1.3$\%$ measured over 1000 runs.  Adding additional sets improves bandwidth but also results in a higher error rate.} 

\section{Side Channel Attack}

We also demonstrate proof of concept side channel attacks.  
The attack primarily uses a spy probe a remote cache and recover a  \textit{memorygram} of the accesses to the cache, which is a collection of cache hits and misses for the different cache sets over time.  Previously, memorygram have been used in cache side channel attacks for website fingerprinting \cite{shusterman2019robust} on CPUs.   This access pattern correlates with the activity on the remote GPU, allowing us to infer information about the applications running on this GPU.  The target of our side channel attack is application fingerprinting on a target GPU in a DGX-1 box. Our side channel attack model is demonstrated in Fig. \ref{fig:threatModel_1}. 
Specifically, the attacker is located on GPU B, and accesses the eviction sets it pre-constructed in its own buffer that is allocated on a remote GPU A. By having the eviction sets for each L2 cache set and measuring the access time on each set, the spy can remotely infer whether the local application has accessed the set (replacing its own data) or not. The memory footprint of applications get recorded over a period of time which reveal the access pattern of the applications on different L2 cache sets. We have demonstrated side channel attack on different genre of applications. Our first side channel attack is on high performance applications and our second attack is on deep learning applications. 

\subsection{Application Fingerprinting using a side channel attack}
We demonstrate a 
side channel where we finger print the remote HPC application based on the memorygram.  Specifically, we pre-train a deep learning network to identify applications based on their memorygram.  This attack can serve as a first step of future attacks where we identify a target application, and then infer information about it.  
This attack 
can be used to identify and reverse engineer the scheduling of applications on a multi-GPU system (simply by spying on all other GPUs in a GPU-box), and identify a target GPUs that are running a specific victim application, and even identify the kernels running on each GPU.


In our proof of concept attack, we used six different applications from NVidia toolkit\cite{cudasample} as our victim applications.\nael{Why not the whole set of applications?} Our application set include common HPC workloads like vectoradd, histogram, blackscholes, matrix multiplication, quasirandom and welshtransform.  

\begin{figure}[ht]
\vspace{-0.1in}
\centering
  \begin{tabular}{lcr}
    \includegraphics[width=.45\linewidth]{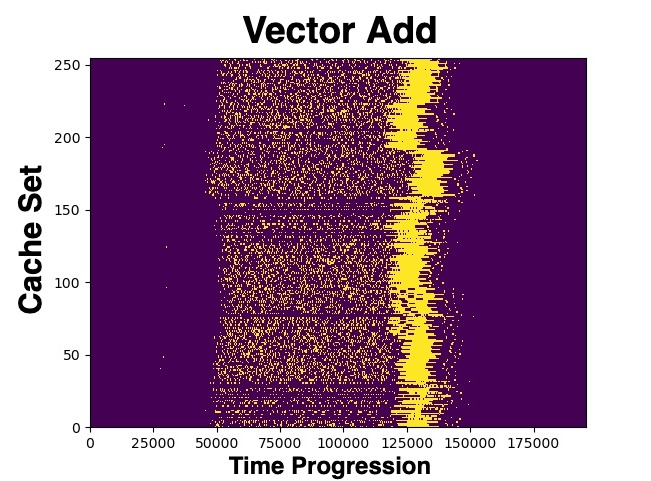} 
    \includegraphics[width=.45\linewidth]{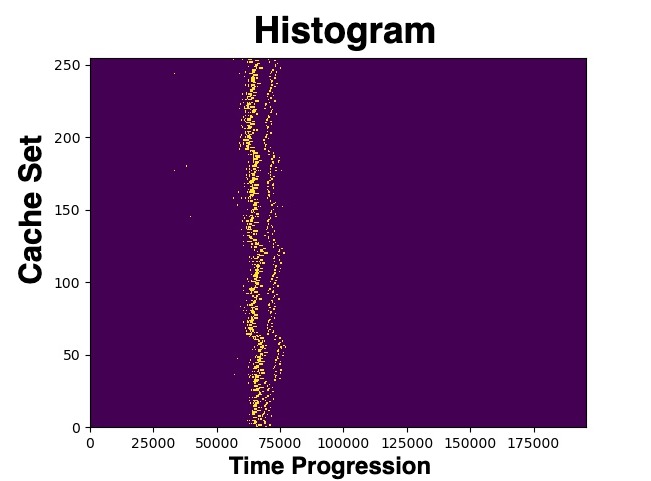} \\
    \includegraphics[width=.45\linewidth]{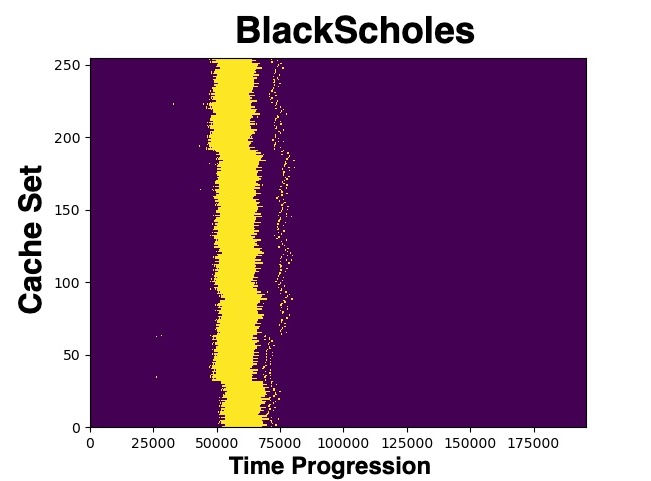} 
    \includegraphics[width=.45\linewidth]{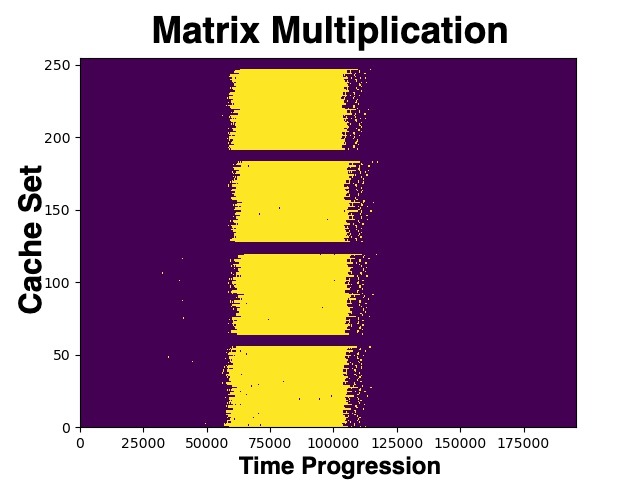} \\
    \includegraphics[width=.45\linewidth]{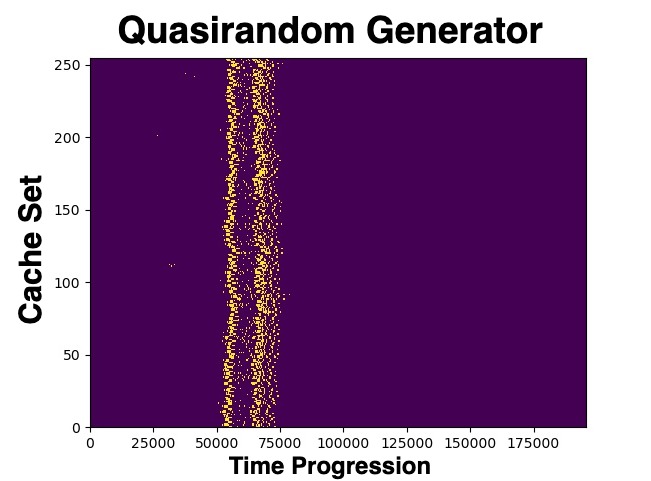} 
    \includegraphics[width=.45\linewidth]{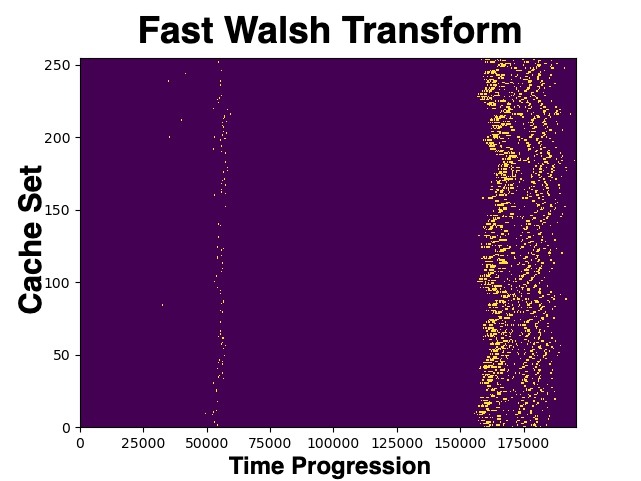}
  \end{tabular}
   \caption{Memorygram of 6 applications}
    \label{fig:memgram}
\end{figure}

Example memorygrams of victim applications are shown in Fig. \ref{fig:memgram}; note that these can be different in each run because the conflict sets hash to arbitrary sets within the cache.  There is some structure, because the hashing preserves page boundaries; that is, the addresses within a single page will hash to consecutive sets in the physical cache.  The X-axis of each image is the execution timeline of the spy application and the Y-axis is the cache set number. The yellow dots represent a cache miss on the L2 each, indicating a likely victim application's access. The image shows the cache misses that occurred on 256 sets of L2 cache. Each victim application leaves a unique memory footprint. 

We train an image classifier to identify the different applications based on input memorygram images (other approaches are possible).  Specifically, we run the attack many times against the different applications to collect $1500$ samples for each application. 
We split the data into training and validation sets of 150 samples each and isolate $1200$ samples as the test or control set. Since there is no class imbalance in the data set, keeping a sufficiently large test set ensures that we evaluate the generalization capabilities with good confidence.

\begin{figure}[htbp]
    \centering
    \includegraphics[height = .5\columnwidth]{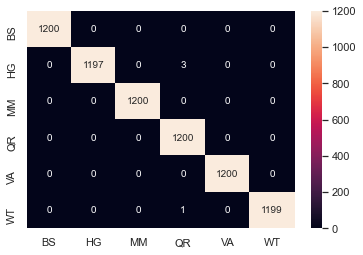}
    \caption{Confusion Matrix. BS (Black Scholes), HG (Histogram), MM (Matrix Multiplication), QR (Quasi Random), VA (Vector Addition), WT (Walsh Transform)}
    \label{fig:confusion_matrix}
    \vspace{-0.1in}
\end{figure}

The classifier achieved an overall accuracy of $99.91\%$ on the test set of $7200$ samples spanning six classes. 
Black Schole, Matrix Multiplication, Quasi Random Generator, and Vector Addition were classified with perfect accuracy score of $100\%$ while Histogram and Welsh Transform scored $99.75\%$ and $99.91\%$ respectively. The confusion matrix depicting the classification results is shown in Figure \ref{fig:confusion_matrix}. We believe the formulation can be readily extended to classify a larger number of applications, and eventually extended to identify specific kernels within an application.  This will enable us to use this attack as a first step to locate the kernels of a long running application and then carry out side channel attacks targeting them individually.

\subsection{Side channel attack on Deep Learning Application}

Machine learning training and inference is perhaps the primary application envisioned for multi-GPU machines. We demonstrate a preliminary side channel targeting extraction of model information from a machine learning model as it executes.  Due to time limitations available for revisions, we demonstrate only the principles of the attack and evidence that it can be successful.

\begin{figure}[ht]
\centering
  \begin{tabular}{lcr}
    \includegraphics[width=\linewidth]{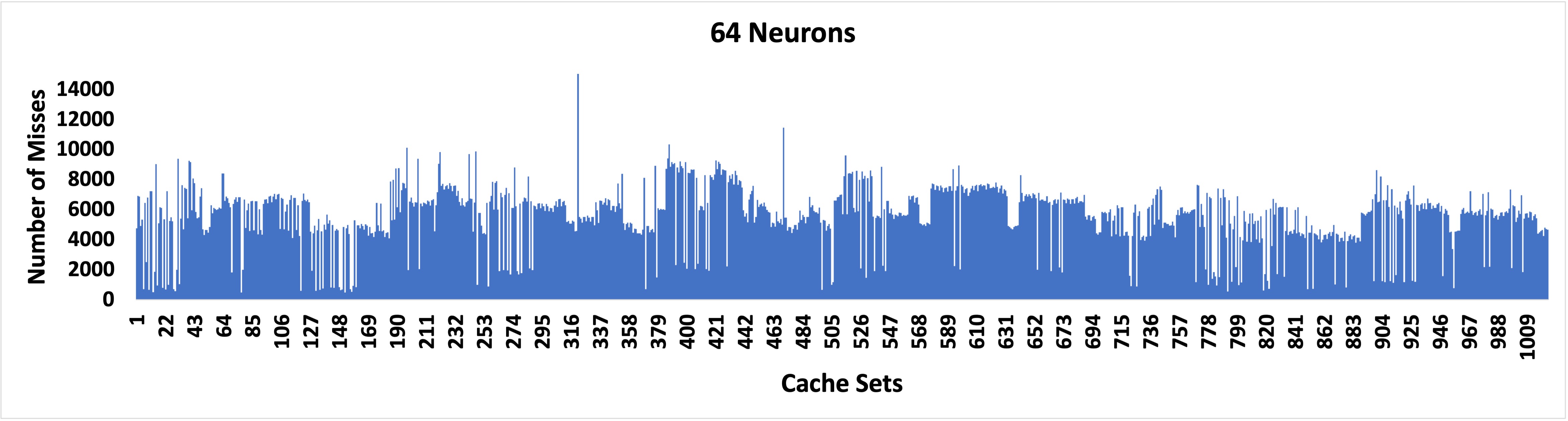}\\ 
    \includegraphics[width=\linewidth]{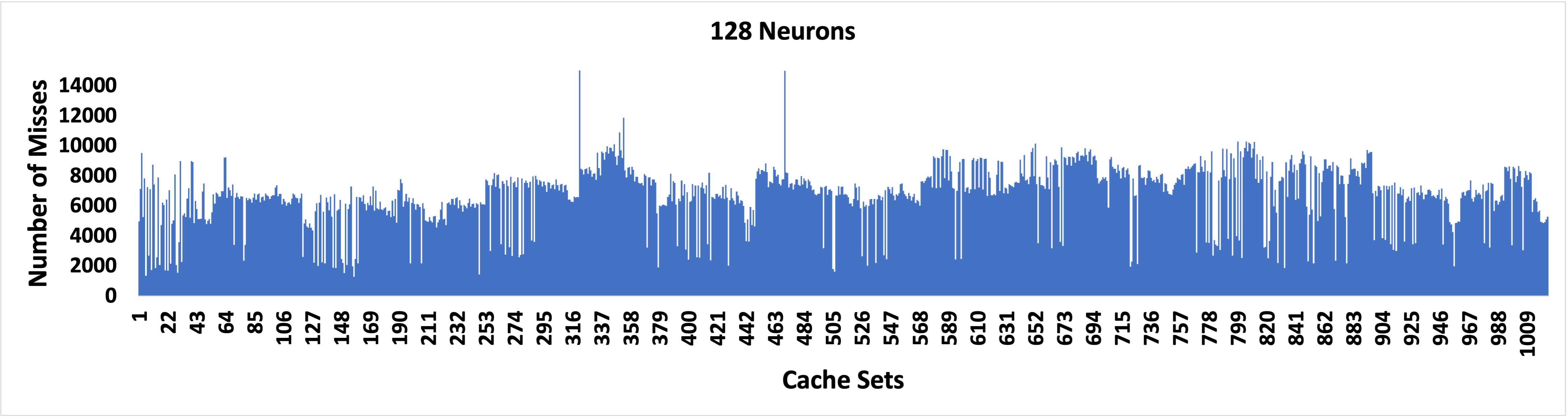} \\
    \includegraphics[width=\linewidth]{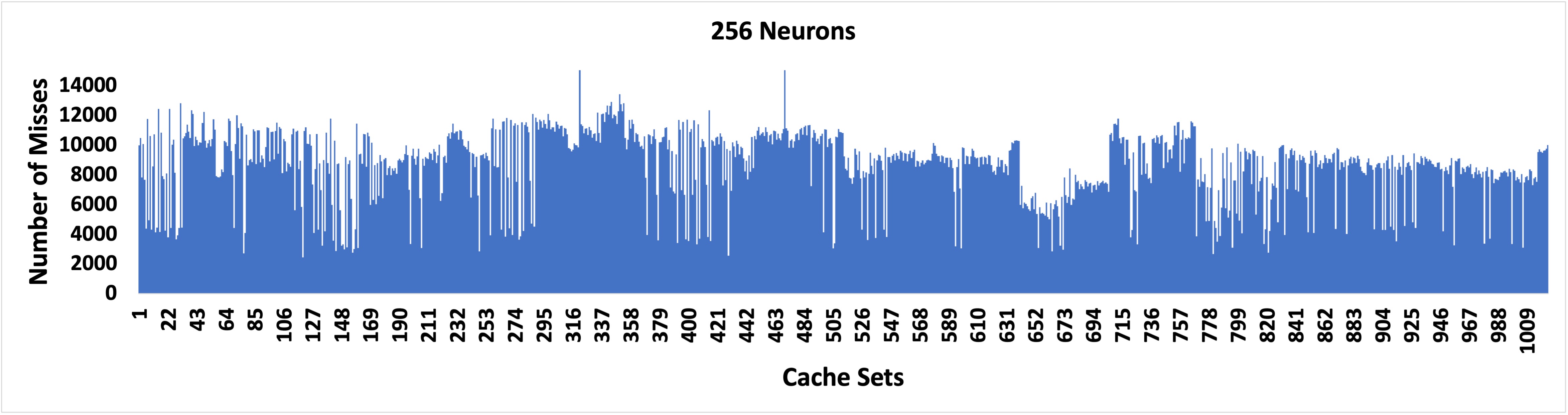}\\
    \includegraphics[width=\linewidth]{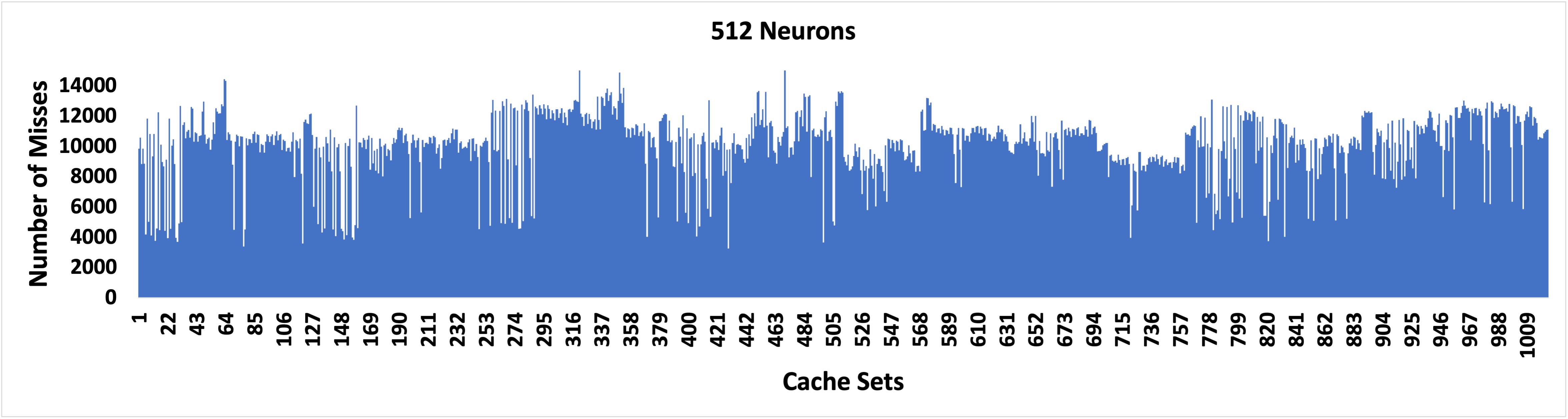}
  \end{tabular}
   \caption{Cache misses per set}
    \label{fig:mlphistogram}
\end{figure}

Our victim attack is a Multi-layer Perceptron (MLP) model with 1 hidden layer built using pytorch \cite{paszke2017automatic}. The application trains the MLP using the MNIST digit recognition data set\cite{deng2012mnist}. 
We have use four different network configurations varying the  number of neurons in the hidden layers; the attack's goal is to identify this number of neurons. 
We monitored 1024 unique L2 cache sets in the remote GPU.  We chose this number to balance sampling coverage and the speed of the attack (how often we can sample each set).
A histogram of the number of misses for each of the monitored cache sets is shown in Fig. {~\ref{fig:mlphistogram}}.  Visually, we can see that the intensity of misses increases as the size of the hidden layer increases, reflecting the additional computations during training.  

\begin{table}
\caption{Average misses over all cache sets}
\centering
\begin{tabular}{ |c|c| } 
\hline
\textbf{Number of Neurons} &  \textbf{Average Number of Misses}\\ 
\hline
64 & 5653\\
\hline
128   & 6846 \\ 
\hline
256 & 8744\\
\hline
512   & 10197\\
\hline
\end{tabular}
\label{tab:avgmiss}
\vspace{-0.20in}
\end{table}

\begin{figure}[htbp!]
    \centering
    \subfloat[Memorygram of MLP with 128 neurons]{
        \label{fig:mlpmemgram128}
        \includegraphics[width=.7\columnwidth]{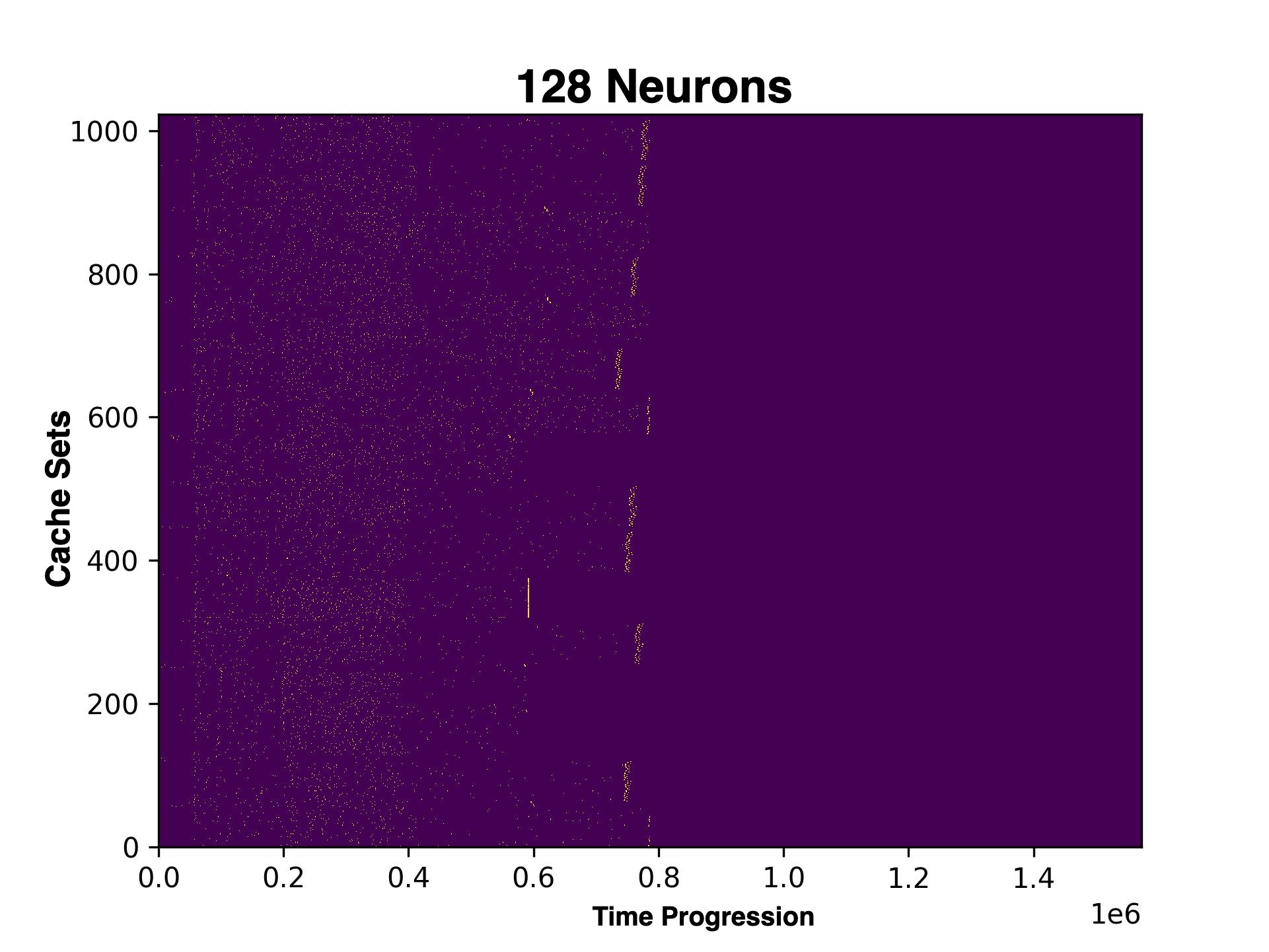}
    }\\
    \subfloat[Memorygram of MLP with 512 neurons]{
        \label{fig:mlpmemgram512}
        \includegraphics[width=.7\columnwidth]{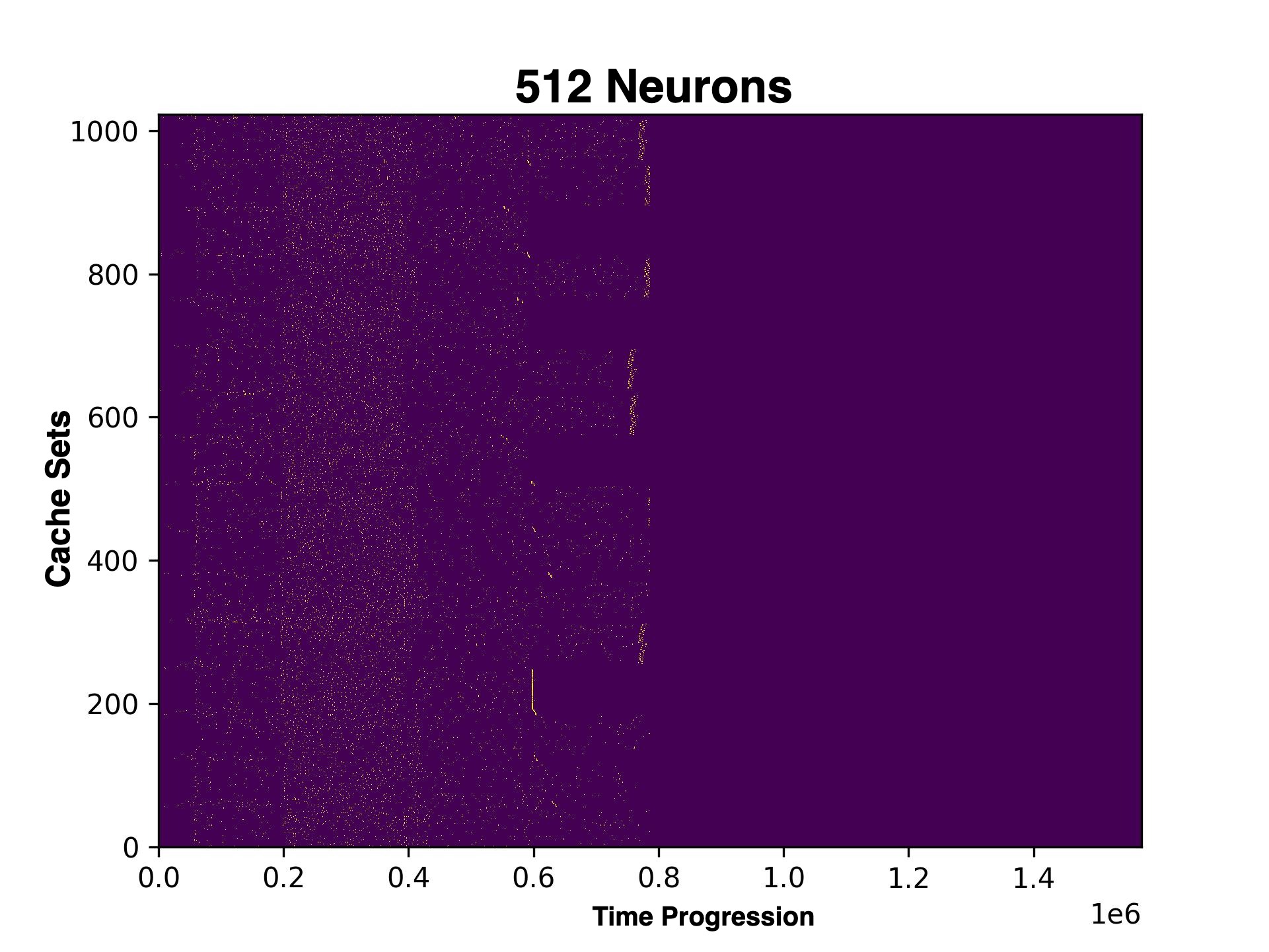}
    }\\
    \caption{Memorygram of the MLP application}
    \label{fig:mlpmemgram}
\end{figure}

\begin{figure}
\vspace{-0.1in}
        \centering
        \includegraphics[width=.7\columnwidth]{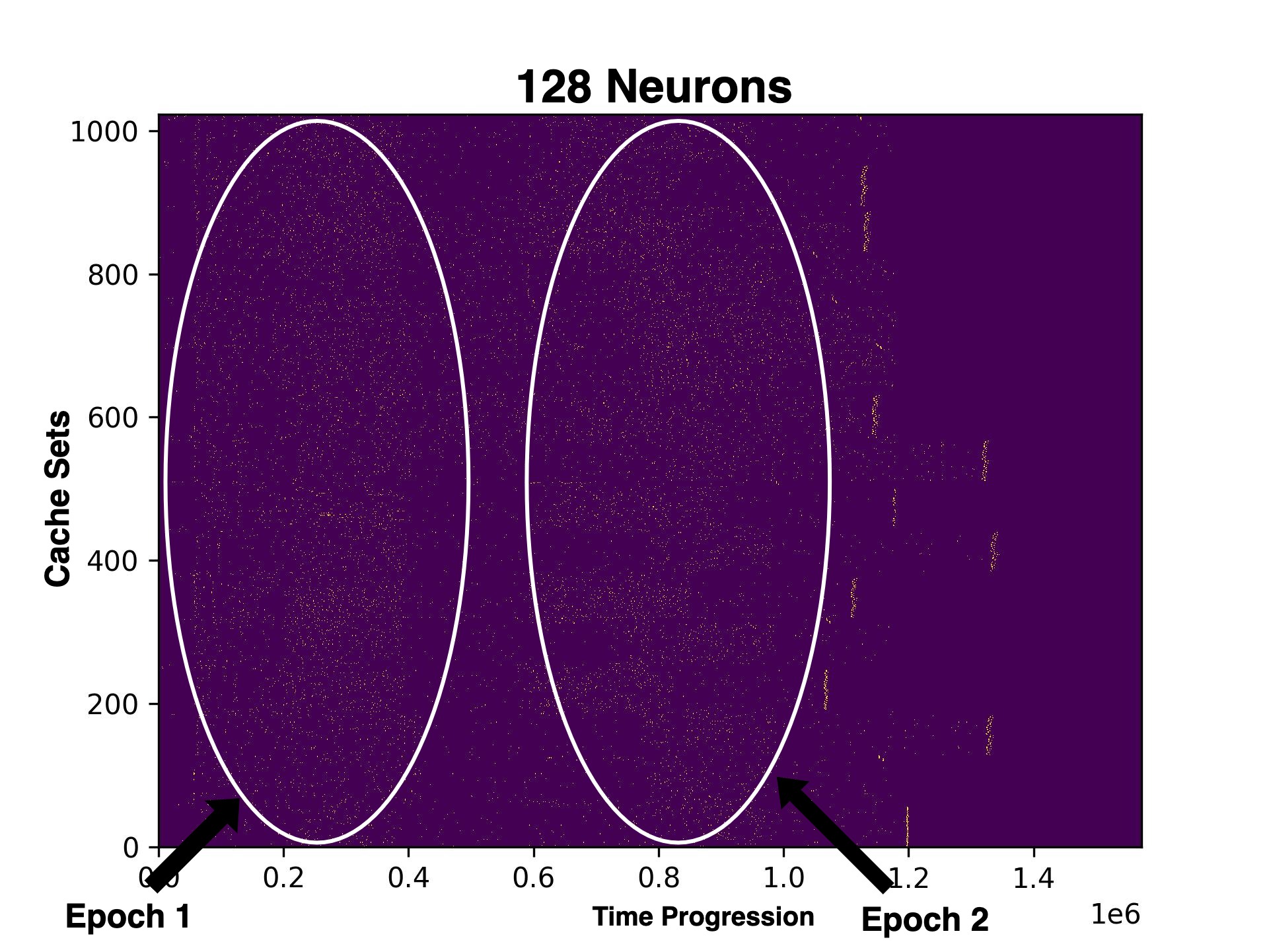}
        \caption{Memorygram for a two-epoch experiment}\label{fig:mlpmemgram128ep2}
\vspace{-0.1in}
\end{figure}

Table {~\ref{tab:avgmiss}} shows the average number of cache set misses; we see  separation which allows us to infer the configuration. Fig. {\ref{fig:mlpmemgram128}} and Fig. {\ref{fig:mlpmemgram512}} show the memorygram of the application with 128 and 512 number of neurons. The memorygram data is  richer, showing the pattern of misses over time, and we believe we can use a classifier on this data to infer more detailed information about the model.   For example, the model was configured to run two epochs in Fig. {~\ref{fig:mlpmemgram128ep2}}  (two full passes through the training set). The number of epochs is a hyperparameter  which we are able to infer visually in Fig. {\ref{fig:mlpmemgram128ep2}}. 

\section{Noise Mitigation}\label{sec:noise}
We developed our attacks in a quiet environment. However, in real scenarios, there will potentially be other concurrent applications running on GPUs, accessing L2 cache and as a result, adding noise to the covert or side channel attacks. 

For mitigating noise, we propose to leverage concurrency limitations of GPUs using similar approaches as prior work~\cite{hoda_micro} to force exclusive execution of spy or trojan on GPUs. Based on leftover policy for GPU multiprogramming, thread blocks of the first process are assigned to different SMs and if there are leftover intra-SM resources for other applications, they can get launched on the same SM concurrently. These resources include shared memory, register, and maximum number of thread block per SM. For example, in covert channel attacks, if we control the resource demand of our trojan on GPU A and spy on GPU B to saturate the intra-SM resources, no other concurrent application can be assigned to those SMs on two GPUs during the covert communication.  Of course, this approach is more difficult for side channels, but it is likely that we would be able to customize a kernel to block out additional noise from the GPU with knowledge of the resources needed by the target victim application.

The attack uses one thread block per SM. However, each thread block can only allocate 32Kb of shared memory on Pascal, which is half the size of the available shared memory per SM.  
To consume the shared memory and block other applications, we launch idle thread blocks to use the leftover shared memory without interfering with the attack (they do not access the global memory during the communication). Therefore, we can ensure the exclusive execution of spy (or trojan) on GPU reducing noise.

\section{Possible Mitigations}\label{sec:Defense}

Defenses against microarchitectural covert and side channel attacks on CPUs and GPUs can potentially apply to cross-GPU attacks with some adaptations. One solution is static or dynamic partitioning of shared resources~\cite{Kong-09,Wang_ISCA'07,domnitser-12,liu-16,GPUGUARD}. For example, Nvidia designed Multi-Instance GPU (MIG) Technology~\cite{mig} in their new generations of discrete GPUs (Ampere). In this design, a single GPU can be securely partitioned into separate GPU instances for multiple users with the isolated paths through the entire memory system; the on-chip crossbar ports, L2 cache banks, memory controllers, and DRAM address busses are all assigned uniquely to an individual instance. However, MIG feature requires privileged access and is not available in Pascal and Volta based DGX machines, still leaving these boxes vulnerable to microarchitectural attacks. 


To minimize the performance overhead of these partitioning-based defense mechanisms, they can only be triggered when contention is detected on a shared resource (similar to the proposed framework in~\cite{GPUGUARD}). In multi-GPU systems, the detection of cross-GPU covert or side channel attacks is possible by monitoring the traffic over NVLinks and access patterns on L2 and memory (accessible through hardware performance counters). In addition, some prior works~\cite{CODA} propose to place the data along with the thread block that accesses it in the same GPU to minimize the remote traffic in multi-GPU systems, and as a result to improve the performance. Although inherent GPU-to-GPU communications can not be completely eliminated in multi-GPU systems, making these cross-GPU data transfers more coarse-grained in normal applications will significantly increase the detection accuracy of high-bandwidth attacks, leading to more efficient defenses.
\section{Related Work}\label{sec:RelatedWork}

With the increasing support of multiprogramming on GPUs in recent years, several works have studied microarchitectural covert and side channel attacks on a single GPU. 

Naghibijouybari et al.~\cite{hoda_micro} characterize contention and construct covert channels on a variety of resources on GPUs, including constant caches, different types of functional units, and memory. Nayak et al.~\cite{Nayak-2021} develop a similar microarchitectural covert channel on GPU’s shared last level translation lookaside buffer(TLB) and Ahn et al.~\cite{ahn-micro21} implement covert channel attacks on shared on-chip interconnect on GPUs. In a completely different environment, Dutta et al.~\cite{dutta_1} developed covert channel attacks between CPU and GPU through shared LLC and ring bus in integrated CPU-GPU systems. \cite{He_aspdac_2020} Conducted a microarchitectural attack on the shared memory in the intel based integrated CPU-GPU systems.

GPU side channel attacks can be categorized in two different threat models: (1) the spy launches the GPU kernel and measures the leakage from the CPU (host) side by exploiting memory coalescing~\cite{jiang-2016, Trident-2021} or shared memory bank conflicts~\cite{jiang-2017} and their correlation with the execution time of GPU kernel, or by collecting hardware performance counters~\cite{Wang-2020}, Electromagnetic traces~\cite{DeepSniffer, gao-2018} and power consumption traces~\cite{mukherjeeside-2015}. Most of these attacks have been implemented to extract the encryption key. (2) the spy is co-located on the GPU with the victim process and measures the contention on the shared resources through hardware performance counters. Through these side channel attacks, the attacker can implement website fingerprinting, inter-keystroke timing attack~\cite{hoda_ccs}, workload fingerprinting ~\cite{Zou-2019,liu-2019}, or Neural Network model extraction attacks~\cite{hoda_ccs, Wei-2020}. 

All of these attacks on discrete GPUs exploit the aggregate measures of contention on GPUs. The attacks that we develop in this paper, are the first Prime+Probe based timing attacks on L2 cache on GPUs, which focus on a single set of cache, providing high-resolution attacks by fine-grained access time measurements. Our attacks also span multiple GPUs in multi-GPU systems, bypassing possible partitioning based mechanisms within a single GPU~\cite{GPUGUARD, mig}.

\section{Concluding Remarks}\label{sec:conclude}

In this paper, we demonstrate for the first time a microarchitectural attack on Multi-GPU systems.  These systems are emerging and increasingly important computational platforms, critical to continuing to scale the performance of important applications such as deep learning.  They are already offered as cloud instances offering opportunities for an attacker to spy on a co-located victim.  We reverse engineer the cache organization and sharing on an Nvidia DGX-1 machine, showing that remote caches can be shared when the attacker allocated memory on the memory banks of the remote GPU.  We reverse engineer the timing properties of both local and remote accesses, as well as the cache replacement policy.  We develop both covert channel and side channel prime-and-probe based attacks across different GPUs.  This attack expands our understanding of the threat model faced by these systems.  For example, we show that defenses designed to protect GPUs against covert and side channel attacks are not set up to prevent these new attacks, which motivates new defenses that can mitigate them.



\newpage
\bibliographystyle{IEEEtranS}
\bibliography{refs}

\end{document}